\documentclass[twocolumn,prb,showpacs,showkeys,amsmath,amssymb]{revtex4}

\pdfoutput=1

\usepackage{graphicx}%
\usepackage{bm}%

\begin{document}

\title{Quantum Key Distribution Using a Triggered \\Quantum Dot Source
Emitting near 1.3~$\mu$m}%

\author{P.~M.~Intallura}
 \altaffiliation[Also at ]{Cavendish Laboratory, University of Cambridge,
 J.~J. Thomson Avenue, Cambridge CB3~0HE, United Kingdom}
 \email{pmi20@cam.ac.uk}
\author{M.~B.~Ward}
\author{O.~Z.~Karimov}
\author{Z.~L.~Yuan}
\author{P.~See}
\author{A.~J.~Shields}
\affiliation{Toshiba Research Europe Limited, Cambridge
Research Laboratory, 208~Cambridge Science Park, Milton Road,
Cambridge CB4~0GZ, United Kingdom}

\author{P.~Atkinson}
\author{D.~A.~Ritchie}
\affiliation{Cavendish Laboratory, University of Cambridge,
J.~J. Thomson Avenue, Cambridge CB3~0HE, United Kingdom}

\date{October 2, 2007}%

\begin{abstract}
We report the distribution of a cryptographic key, secure from
photon number splitting attacks, over 35~km of optical fiber
using single photons from an InAs quantum dot emitting
$\sim$\,1.3~$\mu$m in a pillar microcavity.  Using below
GaAs-bandgap optical excitation, we demonstrate suppression of
multiphoton emission to 10\% of the Poissonian level without
detector dark count subtraction. The source is incorporated
into a phase encoded interferometric scheme implementing the
BB84 protocol for key distribution over standard
telecommunication optical fiber. We show a transmission
distance advantage over that possible with (length-optimized)
uniform intensity weak coherent pulses at 1310~nm in the same
system.
\end{abstract}



\maketitle

The majority of experimental realizations demonstrating quantum
key distribution (QKD) have relied on encoding information onto
weak coherent pulses (WCPs).\cite{Townsend1993EL,Townsend1994EL,Marand1995OL} Due to
the Poissonian nature of laser light, there is a finite
probability of generating two or more photons per pulse from
such a source. This opens up a security threat where an
eavesdropper, Eve, can take advantage of these extra photons by
performing a photon number splitting (PNS)
attack.\cite{Brassard2000PRL} To compensate for this security
loophole, the transmitter, Alice, has to attenuate the signal
by an amount that increases with distance, which reduces the
transmission rate and ultimately limits the maximum secure
transmission distance to the authorized recipient,
Bob.\cite{Gobby2004EL} Decoy-pulse techniques have been
developed to help mitigate the risks associated with
multiphoton pulse emission,\cite{Lo2005PRL,Wang2005PRL} and PNS
secure key distribution distances are now starting to exceed
those achieved with uniform pulse
intensities.\cite{Yin2007Arxive}

A single quantum emitter will exhibit ``anti-bunching'' of the
photon emission times,\cite{Kimble1997PR} such that a regulated
stream of single photons with zero probability of emitting more
than one photon in any given excitation pulse can be expected.
Applying an efficient single quantum emitter in a cryptographic
system would outperform all other methods developed to date.
Several experiments using single photons with wavelengths
compatible with silicon technology have already been used for
QKD. These include using a stream of single-photon pulses
generated by a single nitrogen-vacancy color center in a
diamond\cite{Beveratos2002PRL} and emission from a quantum
dot.\cite{Waks2002Nature} Telecom wavelength QKD has been
achieved by using pairs of photons produced via spontaneous
parametric down conversion.\cite{Trifonov2005JOB,
Soujaeff2007OE}

In this letter we demonstrate QKD using an
optically-excited, triggered single-photon source (SPS)
emitting at a telecom wavelength. Our source, which shows a
ten-fold reduction in multi-photon emission compared to a
laser, has been used to distribute keys secure from the PNS
attack over 35~km along standard optical fiber. By applying a
security analysis for imperfect devices
(GLLP),\cite{Gottesman2004QIC} we demonstrate a transmission
distance advantage compared to the same system configuration
incorporating uniform intensity pulses from a laser source at
the same wavelength.

We have previously demonstrated that a low density of telecom
wavelength dots can be achieved through a bimodal growth
technique\cite{Ward2005APL} and we have used this method to
fabricate a SPS. Our source is based on an InAs quantum dot
embedded at the center of a GaAs spacer in a distributed Bragg
reflector (DBR) pillar microcavity $\sim$\,3~$\mu$m in
diameter. The DBR mirrors consist of alternating layers of GaAs
and Al$_{0.98}$Ga$_{0.02}$As, each with a designed thickness
corresponding to one-quarter optical wavelength. The cavity was
formed with 11 periods on top and 30 periods on the bottom of a
nominally three-optical-wavelength thick spacer layer. The
sample was placed in a custom built confocal microscope which
was cooled in a variable temperature continuous flow liquid
helium cryostat.

\begin{figure}[!t]
\centering\includegraphics[width=\linewidth]{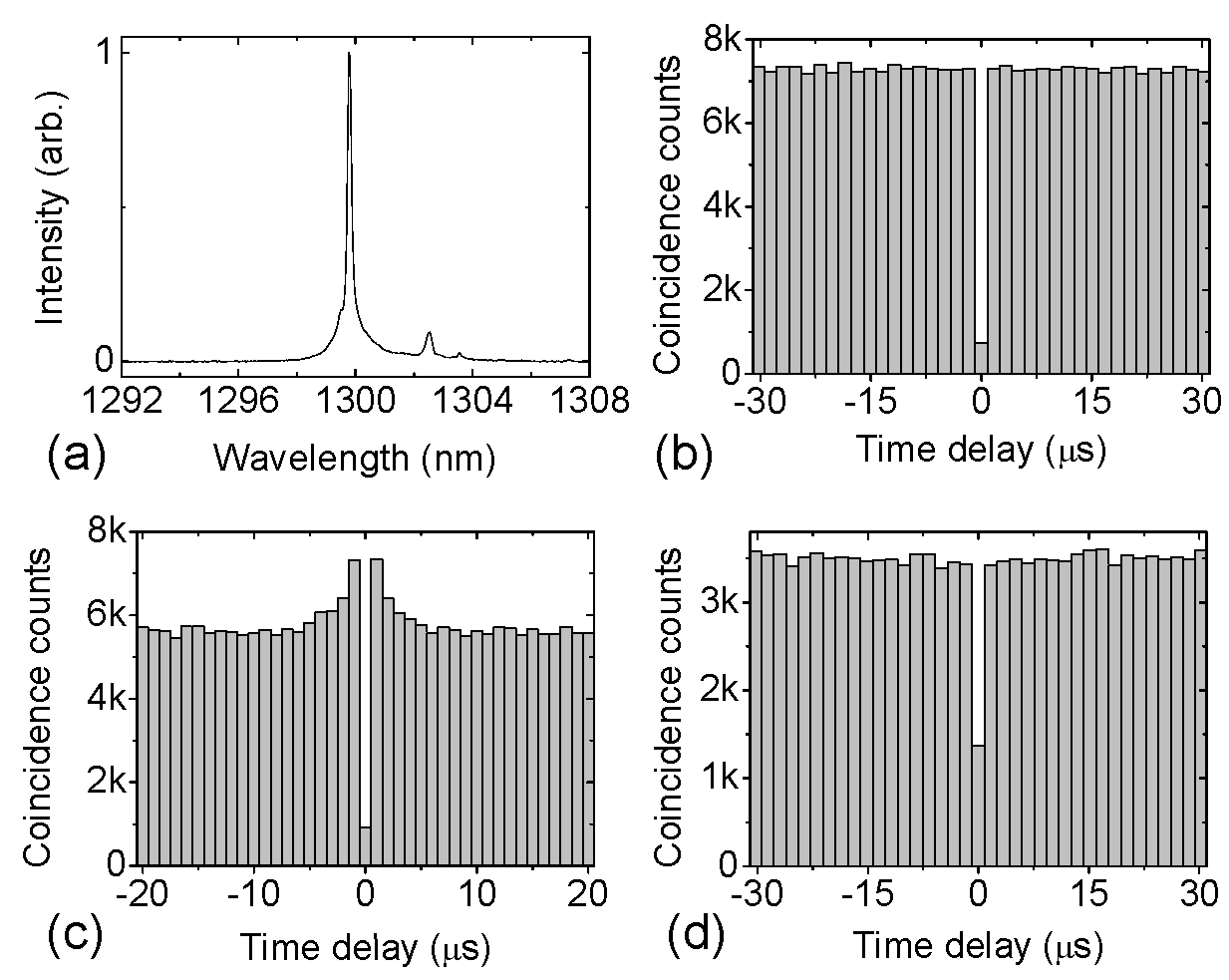}
\caption{\label{fig:CorrCurves} a) Normalized PL spectrum recorded $\sim$\,71~K
before spectral filtering under 1064~nm excitation at saturation power.
Correlation measurements ($\sim$\,71~K) on the spectrally filtered emission line with b)
1064~nm excitation pulsed at 80~MHz at low power ($\sim$\,102~$\mu$W)
with APDs gated at 594~kHz, c) 1064~nm laser excitation at 1~MHz around saturation power
($\sim$\,2.5~$\mu$W at this frequency) with detection at 1~MHz and d) 780~nm
excitation pulsed at 80~MHz near saturation power ($\sim$\,57~$\mu$W)
with detection at 594~kHz. Counts within 150~$\mu$s of a preceding count in the
same detector were rejected to limit the effect of afterpulsing.  Powers quoted refer
to the average power at the input of the microscope.}
\end{figure}

Below GaAs-bandgap laser pulses at 1064 nm were used to trigger photon emission from the quantum dot.  At this wavelength electron-hole pairs are created in highly excited states of the dot.  The telecom wavelength quantum dots used here offer stronger confinement than dots emitting below 1 $\mu$m and several shells are expected to be confined.  The photogenerated carriers are expected to thermalize down to lower energy configurations in time scales shorter than the radiative lifetime. Maximum
single-photon signal from the device was achieved at
$\sim$\,71~K where a charged exciton was on resonance with the
HE$_{11}$ cavity mode [Fig.~\ref{fig:CorrCurves}(a)]. We have
measured suppression of $g^{(2)}(0)=0.102\pm0.004$ as shown in
Fig.~\ref{fig:CorrCurves}(b) without any subtraction of
background emission or detector dark counts. We believe that
this is the lowest measured value of $g^{(2)}(0)$ reported to
date from a quantum dot emitting at a telecom wavelength.

For the purpose of quantum key distribution the source was
operated with excitation intensities around those where the
emission line shows maximum intensity. In this case we measured
$g^{(2)}(0)=0.166\pm0.005$ with 1064~nm optical excitation
[Fig.~\ref{fig:CorrCurves}(c)]. This is markedly lower than for
excitation at 780~nm where a value of
$g^{(2)}(0)=0.392\pm0.011$ was obtained
[Fig.~\ref{fig:CorrCurves}(d)]. With below-band excitation, we measure an efficiency of $\sim\,$5.1$\%$ after coupling into single-mode fiber (SMF28).  This is the highest useful efficiency reported to date at a telecom wavelength and was demonstrated with a simultaneously good $g^{(2)}(0)=0.166$, measured without subtraction of background or detector dark counts.  (The efficiency quoted above was calculated using the correlation measurement count rates.  If the finite measured $g^{(2)}(0)$ is assumed to have arisen from a Poissonian background a single-photon efficiency of 4.6$\%$ would be estimated.) Even higher efficiencies may be achieved by optimizing the coupling of the source to single mode fiber and reducing the roughness of the micropillars to
enhance the cavity quality factor.\cite{Reitzenstein2007APL}

\begin{figure}[!t]
\centering\includegraphics[width=\linewidth]{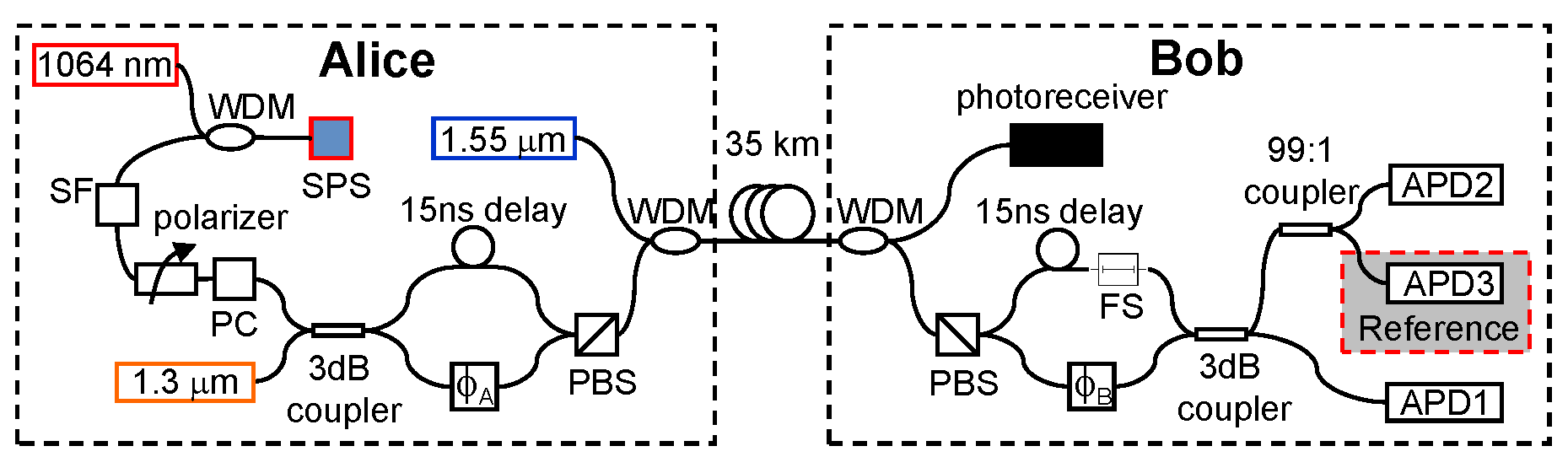}
\caption{\label{fig:QKDSchematic} A schematic diagram
showing the fiber optic quantum key distribution system.
SPS:\ single-photon source; WDM:\ wavelength division multiplexer;
SF:\ spectral filter; PC:\ polarization controller; PBS:\ polarizing beam
splitter/combiner; FS:\ fiber stretcher. Standard telecom fibers are
used in the link between Alice and Bob.}
\end{figure}

\begin{figure}[!t]
\centering\includegraphics[width=\linewidth]{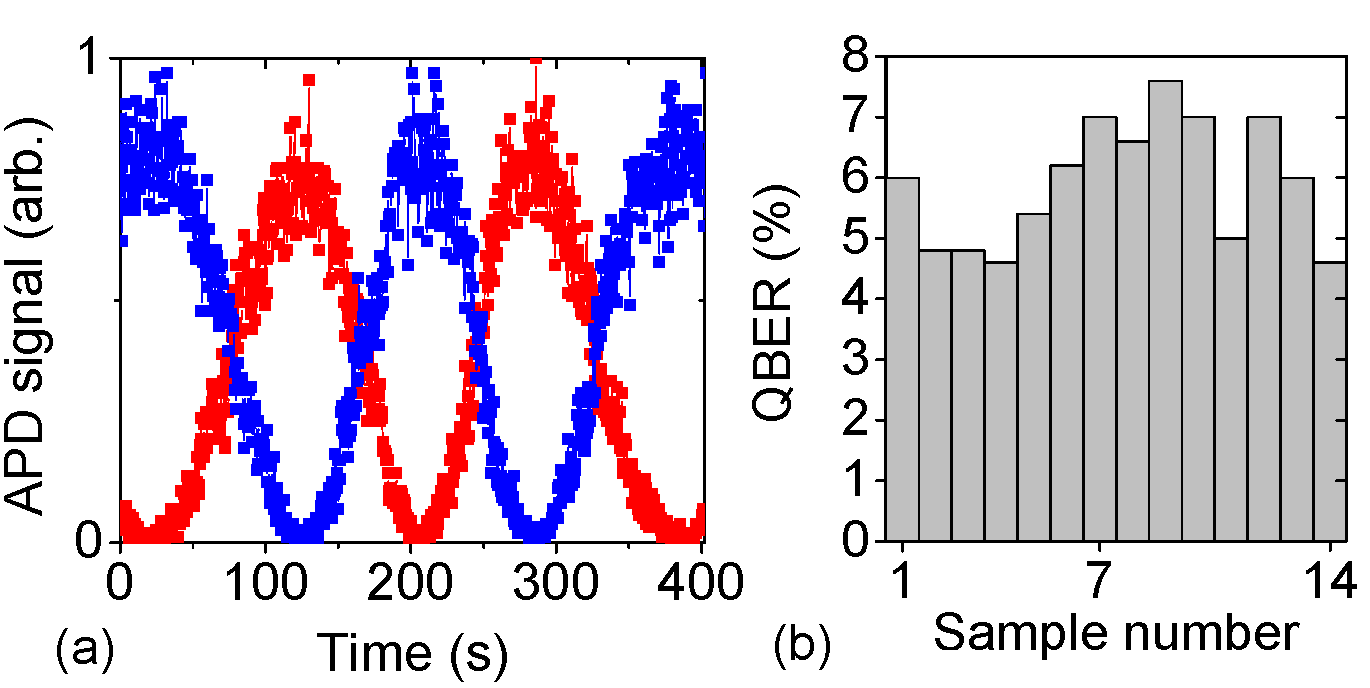}
\caption{\label{fig:QBER} a) Counts as a function of time
on APD1 and APD2 showing a 98.5\% visibility before dark count
correction and b) variation in the QBER over the course of a key transfer
over 35~km of fiber. Each bin represents a sample of 500 sifted bits.}
\end{figure}

Our key distribution system is based upon a time division
Mach-Zender interferometer using phase modulation, as shown in
Fig.~\ref{fig:QKDSchematic}. The system is an extension of that
which our group has used to demonstrate key distribution using
WCPs emitting at 1550~nm.\cite{Gobby2004APL,Yuan2005OE}
However, in this case, we employ our quantum dot source in
place of an attenuated laser and multiplex it with a
1.55~$\mu$m clock laser, which is used as a timing reference.
The source is optically pumped with picosecond-pulsed 1064~nm
laser light from a semiconductor laser diode at a frequency of
1~MHz. A third laser, emitting around the same wavelength as
the source, is introduced to provide feedback to a fiber
stretcher which minimizes any path mismatch in the two
interfering routes.\cite{Yuan2005OE} The BB84
protocol\cite{Bennett1984IEEE} is implemented by
encoding/decoding bit information through phase modulators in
the two interfering paths\cite{Townsend1993EL,Marand1995OL} and
key distribution using the SPS was optimized over 35~km of
optical fiber. InGaAs avalanche photodiodes (APDs) operating in
gated mode at 1~MHz ($\sim$\,165~K) were used to detect the
interfering single photons. The detectors used are around 9\%
efficient at the source wavelength and in the system each gave
a typical count probability of 0.87$\times10^{-6}$ per gate
with the SPS disabled. Of this we attribute a probability of
0.82$\times10^{-6}$ per gate to detector dark counts and the
remainder due to stray counts originating from the lasers. We
estimate a 7.9~dB loss in Alice, a 1.7~dB loss in Bob and use
standard SMF28 telecom fiber with a measured loss of 12.8~dB
through the fiber link. An average visibility of
$\sim$\,99.7$\%$, after dark count correction, was achieved and
is shown in Fig.~\ref{fig:QBER}(a).

A 7~kbit sifted key was distributed with a quantum bit error
rate (QBER) of 5.9\% [see Fig.~\ref{fig:QBER}(b)]. Based on the
GLLP theory for imperfect devices in Gottesman \textit{et
al.},\cite{Gottesman2004QIC} (specifically theorem 6) we can
deduce that the key exchange would have been secure against the
PNS attack given the measured value of $g^{(2)}(0)$ of 0.166.
Eve is assumed to have the full technological capabilities
necessary to access all the information carried by the residual
multiphoton pulses and those pulses contributing to the QBER.
In the security analysis for the SPS, the multiphoton
probability is reduced from $\sim$\,$\mu^{2}/2$ in the WCP case
to $\sim$\,$g^{(2)}(0)\times\mu^{2}/2$, where $\mu$ is the
average number of photons per pulse.

\begin{figure}[!b]
\centering\includegraphics[width=\linewidth]{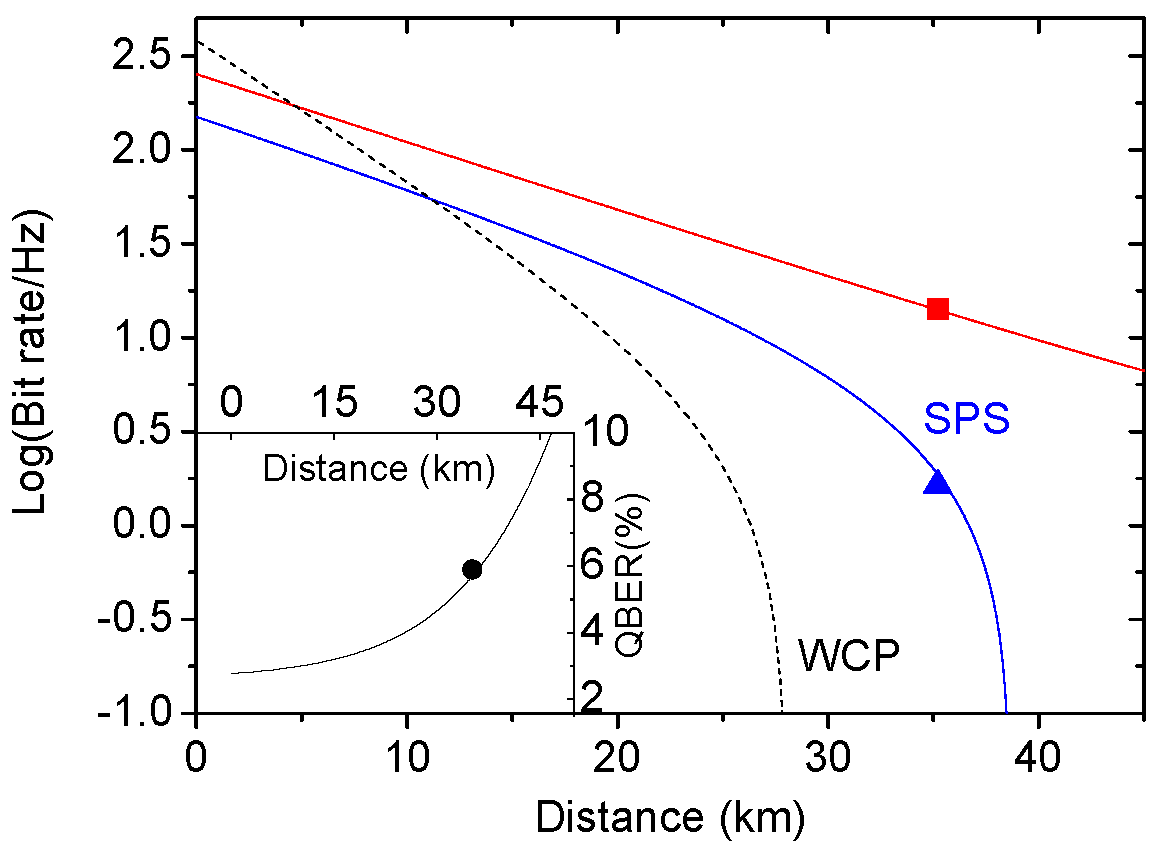}
\caption{\label{fig:GLLP2} Theoretical evaluation
comparing the PNS secure transmission distance of our SPS
(solid blue curve) with an equivalent WCP system (dashed curve).
The solid red curve shows the theoretical sifted bit rate of the SPS. The data
points represent our experimental sifted bit rate (square) and
secure bit rate (triangle). The inset shows the calculated QBER (line) with
the measured value.}
\end{figure}

We now compare the performance of our single-photon BB84 setup
with that which could be achieved using 1310~nm WCPs in the
same system according to GLLP theory. The comparison is made by
taking the same detector gating conditions and assuming the
same dark count probability and detection efficiency. In both
cases the quantum bit errors in the raw key were taken to
consist of half the sum of the dark and stray counts plus 2.6\%
of the raw bit rate due to errors in the phase modulators etc.
The error correction algorithm was taken to use 1.17 times the
number of bits specified in the Shannon
limit.\cite{Brassard1994LNCS,Shannon1948BSTJ} For each fiber
length up to $\sim$\,28~km there exists a range of $\mu$ for
which unconditionally secure QKD using WCPs is possible.
Between the upper bound governed by the PNS attack and a lower
bound governed by the dark (and stray light) count contribution
to the QBER, an optimal laser intensity exists for which the
maximal secure bit rate can be obtained. It is this optimally
attenuated laser performance at 1310~nm that we compare to our
SPS performance in Fig.~\ref{fig:GLLP2}.

Fig.~\ref{fig:GLLP2} shows that key distribution with our dot
source has a quantitative advantage at long distances compared
with that which could be securely achieved using uniform
intensity laser pulses at 1310~nm. We note that for
transmission distances greater than $\sim$\,11~km a higher bit
rate is achieved with the SPS. For short transmission distances
WCP systems are favorable as higher bit rates, due to the
higher photon flux, can be achieved. However, this is an
advantage which could be overcome by improving the performance
of our SPS.

We have demonstrated single-photon quantum key distribution
using a quantum dot source emitting near 1.3~$\mu$m. The source
was excited below the GaAs-bandgap and showed strong
suppression of multi-photon emission to $\sim$\,10\% of the
Poissonian level without detector dark count subtraction. The
below GaAs-bandgap excitation was responsible for the high
quality single-photon emission. We distributed a key secure
against the PNS attack over 35~km. These results highlight the
potential advantage of single-photon QKD for long distance
quantum cryptography through an existing telecom
infrastructure.

\vspace{3mm}

The authors acknowledge partial financial support from DTI and
EPSRC through LINK OSDA Programme QLED, EU through IPs SECOQC
and QAP, and NoE SANDiE.

\raggedright


\begin{references}

\bibitem{Townsend1993EL} P.~D.~Townsend, J.~G.~Rarity, and
    P.~R.~Tapster, Electron.\ Lett.\ {\bf 29}, 634 (1993).
\bibitem{Townsend1994EL} P.~D.~Townsend, Electron.\ Lett.\ {\bf
    30}, 809 (1994).
\bibitem{Marand1995OL} C.~Marand, and P.~D.~Townsend, Opt.\
    Lett.\ {\bf 20}, 1695 (1995).
\bibitem{Brassard2000PRL} G.~Brassard, N.~L\"{u}tkenhaus,
    T.~Mor, and B.~C.~Sanders, Phys.\ Rev.\ Lett.\ {\bf 85},
    1330 (2000).
\bibitem{Gobby2004EL} C.~Gobby, Z.~L.~Yuan, and A.~J.~Shields,
    Electron.\ Lett.\ {\bf 40}, 1603 (2004).
\bibitem{Lo2005PRL} H.-K.~Lo, X.~Ma, and K.~Chen, Phys.\ Rev.\
    Lett.\ {\bf 94}, 230504 (2005).
\bibitem{Wang2005PRL} X.-B.~Wang, Phys.\ Rev.\ Lett.\ {\bf 94},
    230503 (2005).
\bibitem{Yin2007Arxive} Z.-Q.~Yin, Z.-F.~Han, W.~Chen,
    F.-X.~Xu, Q.-L.~Wu, and G.-C.~Guo,
    arXiv:0704.2941v2~[quant-ph] (2007).
\bibitem{Kimble1997PR} H.~J.~Kimble, M.~Dagenais, and
    L.~Mandel, Phys.\ Rev.\ Lett.\ {\bf 39}, 691 (1977).
\bibitem{Beveratos2002PRL} A.~Beveratos, R.~Brouri, T.~Gacoin,
    A.~Villing, J.-P.~Poizat, and P.~Grangier, Phys.\ Rev.\
    Lett.\ {\bf 89}, 187901 (2002).
\bibitem{Waks2002Nature} E.~Waks, K.~Inoue, C.~Santori,
    D.~Fattal, J.~Vu\v{c}kovi{\'{c}}, G.~S.~Solomon, and
    Y.~Yamamoto, Nature {\bf 420}, 762 (2002).
\bibitem{Trifonov2005JOB} A.~Trifonov, and A.~Zavriyev, J.\
    Opt.\ B\ {\bf 7}, S772 (2005).
\bibitem{Soujaeff2007OE} A.~Soujaeff, T.~Nishioka, T.~Hasegawa,
    S.~Takeuchi, T.~Tsurumaru, K.~Sasaki and M.~Matsui, Opt.\
    Express {\bf 15}, 726 (2007).
\bibitem{Gottesman2004QIC} D.~Gottesman, H.-K.~Lo,
    N.~L\"{u}tkenhaus, and J.~Preskill, Quantum Information and
    Computation {\bf 4}, 325 (2004).
\bibitem{Ward2005APL} M.~B.~Ward, O.~Z.~Karimov, D.~C.~Unitt,
    Z.~L.~Yuan, P.~See, D.~G.~Gevaux, A.~J.~Shields,
    P.~Atkinson, and D.~A.~Ritchie, Appl.\ Phys.\ Lett.\ {\bf
    86}, 201111 (2005).
\bibitem{Reitzenstein2007APL} S.~Reitzenstein, C.~Hofmann,
    A.~Gorbunov, M.~Strau{\ss}, S.~H.~Kwon, C.~Schneider,
    A.~L\"{o}ffler, S.~H\"{o}fling, M.~Kamp, and A.~Forchel,
    Appl.\ Phys.\ Lett.\ {\bf 90}, 251109 (2007).
\bibitem{Gobby2004APL} C.~Gobby, Z.~L.~Yuan, and A.~J.~Shields,
    Appl.\ Phys.\ Lett.\ {\bf 84}, 3762 (2004).
\bibitem{Yuan2005OE} Z.~L.~Yuan, and A.~J.~Shields, Opt.\
    Express {\bf 13}, 660 (2005).
\bibitem{Bennett1984IEEE} C.~H.~Bennett and G.~Brassard,
    \textit{Proceedings of IEEE International Conference on Computers,
    Systems and Signal Processing}, (IEEE, New York, 1984), p.~175.
\bibitem{Brassard1994LNCS} G.~Brassard, and L.~Savail, Lect.\
    Note\ Comput.\ Sci.\ {\bf 765}, 410 (1994).
\bibitem{Shannon1948BSTJ} C.~Shannon, Bell Syst.\ Tech.\ J.\
    {\bf 27}, 379 (1948).
\end{references}
\end{document}